# Physical properties of the non-centrosymmetric superconductor $Nb_{0.18}Re_{0.82}$

A.B. Karki, Y.M. Xiong, N. Haldolaarachchige, S. Stadler, I. Vekhter, P.W. Adams, and D.P. Young

*Department of Physics and Astronomy, Louisiana State University, Baton Rouge, Louisiana, 70803, USA*

W. A. Phelan and Julia Y. Chan

*Department of Chemistry, Louisiana State University, Baton Rouge, Louisiana, 70803, USA*

We report the synthesis and measurements of magnetic, transport, and thermal properties of polycrystalline $Nb_{0.18}Re_{0.82}$, which has a superconducting transition at $T_c \sim 8.8$ K. The non-centrosymmetric α-Mn structure of the compound is confirmed by X-ray diffraction. Using the measured values for the lower critical field $H_{c1}$, upper critical field $H_{c2}$, and the specific heat $C$, we estimate the thermodynamic critical field $H_c(0)$, coherence length $\xi(0)$, penetration depth $\lambda(0)$, and the Ginzburg-Landau parameter $\kappa(0)$. The specific heat jump at $T_c$, $\Delta C/\gamma T_c = 1.86$, suggests that $Nb_{0.18}Re_{0.82}$ is moderately coupled superconductor. Below $T_c$ the electronic specific heat decays exponentially, suggesting that the gap is isotropic. Our data suggests that the triplet admixture is weak in the polycrystalline form of compound. However, the estimated value of the upper critical field $H_{c2}(0)$ is close to the calculated Pauli limit indicating the need for single crystal measurements.

## I. INTRODUCTION

The topic superconductivity in systems lacking spatial inversion symmetry has undergone a resurgence since the discovery of superconductivity in $CePt_3Si$.[1] Much of the research effort has been aimed at understanding the effects of inversion symmetry on superconducting order parameter. In noncentrosymmetric superconductors, the antisymmetric (in the electron momentum) spin-orbit coupling (ASOC) breaks parity symmetry, so that parity is no longer a good quantum number. Instead of spin, the quasiparticle states are classified by their helicity, the relative orientation of their spin to the crystal momentum. Consequently, the Fermi surface is split into sheets of different helicity by the spin-orbit interaction. Superconducting pairing on a given helicity sheet is comprised of a mixture of spin singlet and triplet eigenstates, and may exhibit unusual magnetic properties and/or display nodes in the energy gap, depending on the structure and the strength of the triplet component.[2-6] In contrast to the triplet *p*-wave superconductors proposed in centrosymmetric systems, such as $Sr_2RuO_4$, the lack of inversion symmetry makes it possible for the triplet component to exist in a fully spatially-symmetric paired state, so that one can speak, for example, of a triplet *s*-wave superconductor.[3] Similarly, nodes in the superconducting gap may exist even when the pair wave function exhibits the full spatial symmetry of the crystal.[2-7]

Noncentrosymmetric heavy fermion superconductors have attracted most of the recent attention due to the unconventional behavior observed in these strongly correlated electron compounds.[1,8-12] However, transition metal compounds such as $Li_2(Pd, Pt)_xB$, $Mg_{10}Ir_{19}B_{16}$, $(Rh, Ir)Ga_9$, and $Mo_3Al_2C$ are more straightforward systems for exploring the ramifications of inversion symmetry breaking. These systems generally exhibit weak electron correlation effects, making it simpler to isolate the salient features of ASOC-mediated superconductivity.[13-22]

In this paper we focus on the intermetallic binary compound $Nb_{0.18}Re_{0.82}$ which does not possess spatial inversion symmetry and displays bulk superconductivity below 9 K. This structural phase ($\chi$-phase) of the Nb-Re system is isomorphous with α-Mn. It is cubic, with space group I$\bar{4}$3m, and contains 58 atoms in its unit cell occupying four crystallographically distinct sites. Savitskii *et. al.* reported that $Nb_{0.18}Re_{0.82}$ with a lattice constant of 9.641 Å exhibits superconductivity at a moderately high temperature near 9.7 K.[23,24] However there have been no comprehensive reports on the electronic structure nor the low temperature electronic properties of this material.

## II. EXPERIMENT

A polycrystalline sample (0.3 g) of $Nb_{0.18}Re_{0.82}$ was prepared by melting stoichiometric amounts of high purity niobium powder (99.95 % Alfa Aesar) and rhenium shot (99.999 % Alfa Aesar) under flowing ultra-high-purity argon gas in an arc-melter using a tungsten electrode and a water-cooled copper hearth. A Zr button was utilized as an oxygen getter and melted before each synthesis. First, Nb and Re buttons were made by arc melting the Nb and Re powder alone. The Nb and Re buttons were then melted together to make a single button of $Nb_{0.18}Re_{0.82}$. The sample was inverted and re-melted several times to ensure homogeneous mixing of the constituent elements. The sample formed a hard pellet with a negligible mass loss. We made more than 5 samples with different concentrations of Nb and Re in order to synthesize the stoichiometric compound and to check the effects of off-stoichiometry on the superconducting transition temperature $T_c$.

X-ray diffraction analysis on well-ground powder of a portion of the sample was carried out on a Bruker Advance D8 powder diffractometer equipped with Cu $K_\alpha$ radiation ($\lambda$ = 1.54056 Å) or on a Scintag XDS2000 powder X-ray diffractometer. The system employs a theta-2theta geometry and Cu Kα X-ray source. Data were collected from $2\theta = 20^o$ to $70^o$ with a constant scan speed of 2º min$^{-1}$ at room temperature. Elemental analysis of the sample was performed using wavelength dispersive X-ray spectroscopy (WDS) with a JEOL JXA-733 SuperProbe Electron Probe Microanalyzer (EPMA).

The electrical resistivity was measured using a standard 4-probe ac technique at 27 Hz with an excitation current of 3 mA, in which small diameter Pt wires were attached to the sample using a conductive epoxy (Epotek H20E). Data were collected from 1.8 to 290 K and in magnetic fields up to 9 T using a Quantum Design, Physical Property Measurement System (PPMS). The specific heat was measured in the PPMS using a time-relaxation method between 2 and 20 K at 0 and 9 T. Magnetic susceptibility was measured in a constant magnetic field of 30 Oe; the sample was cooled down to 3 K in zero field, and then magnetic field was applied, followed by heating to 10 K [zero-field cooled (ZFC)] and then cooled down again to 3 K [field cooled (FC)] in the PPMS.

## III. RESULTS AND DISCUSSIONS

The X-ray powder diffraction pattern of an arc-melted sample of $Nb_{0.18}Re_{0.82}$ is shown in Fig. 1. The pattern indicates the sample was single phase and the cubic cell parameter a = 9.6293 Å was obtained. A schematic view of the crystal structure of $Nb_{0.18}Re_{0.82}$ is presented in the inset of Figure 1. It has the α-Mn structure. The primitive cubic Bravais lattice adopts the space group $I4$-$3m$ (No. 217) with Pearson symbol $cI$58 and lacks a center of inversion symmetry.[24] The WDS elemental analysis of the sample of $Nb_{0.18}Re_{0.82}$ confirmed the atomic ratio of 18.5 to 81.5 on average.

The temperature dependence of the resistivity of $Nb_{0.18}Re_{0.82}$ between 10 K and 290 K is shown in Fig. 2. The resistivity is metallic, and one observes an inflection point in $\rho(T)$ at ~60 K. The normal-state resistivity is fairly low (~250 $\mu\Omega$ cm at room temperature), although it is of polycrystalline form. However, the residual resistivity ratio ($\rho_{290 K} / \rho_{10 K} = 1.35$) is small, suggesting the transport in the polycrystalline sample is dominated disorder, and perhaps small amounts of impurities. We have fit the low temperature resistivity data to a power law,

$$\rho = \rho_0 + AT^{\alpha}, \qquad (1)$$

with $\alpha = 2$, the residual resistivity $\rho_0 = 189$ $\mu\Omega$-cm and the coefficient $A = 0.007$ $\mu\Omega$ cm / K$^2$. The fit is shown as a solid line in Fig. 2, and it describes the data well between 10 and 50 K. The value of $\rho_0$ is large most likely due to the polycrystalline nature of the sample, but the quality of the fit is quite good, as indicated in the inset of Fig. 3. The value of $A$ is sometimes taken as a measure of the degree of electron correlations in the material, and its value suggests $Nb_{0.18}Re_{0.82}$ is a weakly correlated electron system. The inset of Fig. 2 shows that the resistivity varies as $T^2$ over the above temperature range, a clear Fermi-liquid regime is observed. In Fig. 3 we show the resistivity over the entire range of temperature from 3 to 290 K, with a sharp superconducting transition at 8.75 K. The 90% −10% transition width is less than 0.1 K.

The magnetic characterization of the superconducting transition at 15 Oe is shown in Fig. 4. A large diamagnetic superconducting signal is observed near 8 K, which is slightly below the resistive transition temperature. The superconducting volume fraction of the arc-melted (nearly) spherical button was ~ 100 %, after applying the demagnetization factor to the ZFC data at 3 K. As shown in the figure, the sample has reached a full Meissner state below 6 K. The magnetization vs. temperature data over the range 10 to 70 K at 1000 Oe (not shown here) shows temperature-independent Pauli-paramagnetic behavior with a very small moment ($\chi = 6.99 \times 10^{-4}$ cm$^3$/mol).

The magnetization vs. magnetic field over a range of temperatures below $T_c$ is presented in the inset of Fig. 5. For the analysis, the point of deviation from the full Meissner effect of the data curve was chosen as the lower critical field at each temperature. As expected, $\mu_0 H_{c1}(T)$ varied as $T^2$ in accord with:

$$H_{c1}(T) = H_{c1}(0)[1-(T/T_c)^2]. \qquad (2)$$

A least-squares fit to the data in Fig. 5 using equation 2 yielded $\mu_0 H_{c1}(0) = 0.0055$ T.

The resistive transition was measured in a variety of applied fields for the same sample, as shown in the inset of Fig. 6. As the field was increased, the superconducting transition became broader, and $T_c$ shifted to lower temperature. The mid-point of the resistivity transition was chosen as the transition temperature at each applied magnetic field. It should be noted that a transition to a zero resistance state was obtained even at 9 T and above 3 K, indicating a high upper critical field. In the main panel of Fig. 6, the upper critical field is plotted as a function of temperature. The variation in $H_{c2}$ with temperature is nearly linear with a negative slope, and it does not show any kind of saturation for fields as high as 9 T.

We estimate the orbital upper critical field, $\mu_0 H_{c2}(0)$ from the Werthamer-Helfand-Hohenberg (WHH)[25] expression,

$$\mu_0 H_{c2}(T) = -0.693\, \mu_0\, (dH_{c2}/dT)_{T=T_c}\, T_c\,. \qquad (3)$$

Our evaluation of the slope yeilds, $\mu_0 dH_{c2}/dT \approx -2.318$ T / K and using $T_c = 8.75$ K, we find $\mu_0 H_{c2}(0) = 14.04$ T. A complementary estimate of the upper critical field from the empirical formula

$$H_{c2}(T) = H_{c2}(0)(1-t^2)/(1+t^2), \qquad (4)$$

where $t = T/T_c$, gives $\mu_0 H_{c2}(0) = 17.3$ T, as shown in Fig. 6 by the solid line representing the best fit to the experimental data. This extrapolated value is slightly larger than that obtained from the WHH theory.

The characterization of the superconducting transition by specific heat measurements is shown in Fig. 7. The main panel shows the variation of $C/T$ with $T^2$ at lower temperature and zero field. The data above 8.9 K provide an extrapolation of the normal-state behavior to the $T = 0$ limit, and allow the determination of the Sommerfeld constant ($\gamma$) from the fit $C(T)/T = \gamma + \beta T^2$. The extrapolation produced $\gamma = 53.5$ mJ mol$^{-1}$K$^{-2}$ and $\beta = 2.05$ mJ mol$^{-1}$K$^{-4}$. The value of $\gamma$ implies that Nb$_{0.18}$Re$_{0.82}$ has a moderately enhanced density of states. In a simple Debye model for the phonon contribution to the specific heat, the $\beta$ coefficient is related to the Debye temperature $\Theta_D$ through $\beta = N(12/5)\pi^4 R \Theta_D^{-3}$, where $R = 8.314$ J mol$^{-1}$ K$^{-1}$. Using $N = 1$ for Nb$_{0.18}$Re$_{0.82}$ gives a Debye temperature of $\Theta_D = 98.98$ K. In the inset of Fig. 7, $C/T$ is plotted as a function of $T$ showing a detail of the specific heat jump at the thermodynamic transition. The midpoint of the superconducting transition temperature is ~8.7 K, which is close to the value from the resistivity and magnetic measurements. It is known that the ratio $\Delta C/\gamma T_c$ can be used to measure the strength of the electron coupling.[26] The specific heat jump ($\Delta C/T_c$) for $Nb_{0.18}Re_{0.82}$ is 100 mJ mol$^{-1}$K$^{-2}$, which sets the value of $\Delta C/\gamma T_c$ to 1.86, which is larger than the theoretical value of 1.43 for conventional BCS superconductor. This suggests a moderately enhanced electron-phonon coupling. The coupling strength is not strong even though both the Debye and transition temperature are comparable to that of lead.

The electronic specific heat ($C_e$) below the superconducting transition temperature ($T_c$) is calculated by subtracting the phonon contribution $C_l$ from the total specific heat $C$, see Fig. 8. To evaluate characteristics of the superconducting gap function, $C_e$ was analyzed by fitting it to the forms $e^{-b/T}$, $T^2$, and $T^3$, the expected temperature dependencies for an isotropic, and gaps with line, and linear point nodes, respectively. The data follows the exponential fit $C_e = a\, e^{-b/T}$ fairly well over the entire temperature range, while the quadratic fit is poor. A fit to the data using $T^3$, while better than the quadratic form, is a worse approximation to the data than the exponential fit at very low temperatures, the most important temperature region for the conclusions about the gap shape. The above analysis suggests that the best fit is provided by a fully-gapped model. The data do comply with entropy conservation, as $\int_0^{Tc} \frac{C_e}{T} dT = \gamma_n T_c$. An s-wave BCS model of the entire $C_{es}$ data gives the estimated value of the superconducting gap energy $\Delta(0)$ from the relation $\Delta(0)/k_B = 1.83 T_c$ to be $21.96 \times 10^{-23}$ J or $2\Delta(0) = 3.67 k_B T_c = 31.9$ K. The enhanced value of this ratio is comparable to the energy gap in $Li_2Pd_3B$ (29.6 K) and $Mo_3Al_2C$ (36.18 K). It is higher than in other noncentrosymmetric superconductors, such as $Mg_{10}Ir_{19}B_{16}$ (16 K) and $Li_2Pt_3B$ (7.7 K).[17,18,,27]

An estimation of the strength of the electron-phonon coupling can be derived from the McMillan formula. McMillan's model contains the dimensionless electron phonon coupling constant $\lambda_{ep}$, which is – in terms of the Eliashberg theory- related to the phonon spectrum and the density of states. This parameter $\lambda_{ep}$ represents the attractive interaction, while the second parameter $\mu^*$ is introduced to account for the screened Coulomb repulsion. We follow ref. [37] and set $\mu^* = 0.13$. The modified McMillian formula: $\lambda_{ep} = [1.04 + \mu^* \ln(\omega_{ln}/1.2T_c)] / [(1-0.62\mu^*) \times \ln(\omega_{ln}/1.2T_c) - 1.04]$, where $\omega_{ln}$ is a logarithmic averaged phonon frequency, yields a value of $\lambda_{ep} = 0.73$. This value of $\lambda_{ep}$ agrees with classifying $Nb_{0.18}Re_{0.82}$ as a moderately-coupled superconductor, as suggested by $2\Delta_0/k_B T_c$ and $\Delta C/\gamma T_c$. Here $\omega_{ln}$ was determined to be 264 K from $2\Delta_0/k_B T_c = 3.53[1+12.5(T_c/\omega_{ln})^2 \ln(\omega_{ln}/2T_c)]$. This value of $\omega_{ln}$ is high compared to Debye temperature obtained from the estimate of the heat capacity above. We speculated that, since the number of atoms per unit cell is large, in this material there may exit acoustic modes with low velocity that contribute significantly to C/T but do not substantially influence superconductivity.[28, 29]

The value of $\gamma$ corresponds to the electronic density of sates at the Fermi energy $N(E_F)$ of 12.7 states /(eV f.u.) (f.u. stands for formula unit) estimated from the relation,[30]

$\gamma = \pi^2/3\, k_B^2 N(E_F)(1+\lambda_{ep}).$ (5)

A straight forward calculation of the condensation energy from the relation, $U(0) = \frac{1}{2} \Delta^2(0)N(E_F)$, produces a value of , $U(0) = 1037.2$ mJ/mol.

Assuming that the diamagnetic and van Vleck contributions to the magnetic susceptibility are small we consider $\chi = 6.99 \times 10^{-4}$ cm$^3$/mol as the spin susceptibility and estimate the Wilson ratio $R_w = \pi^2 k_B^2 \chi_{spin} / (3\mu_B^2 \gamma) = 0.94$ which is close to 1- the value for free a electron. When normalizing $A$, the coefficient of the quadratic resistivity term, by the effective mass $\gamma$, the Kadowaki-Woods ratio $A/\gamma^2$ was obtained to be 13a$_0$, where a$_0 = 10^{-5}$ $\mu\Omega$ cm/ (mJ/ mol K)$^2$. The smaller values of both $R_w$ and $A/\gamma^2$ confirm that Nb$_{0.18}$Re$_{0.82}$ is weakly correlated electron system.[31-33]

If we assume the upper critical field to be purely orbital, the superconducting coherence length can be estimated using $H_{c2}(0) = \Phi_0/2\pi\xi(0)^2$, where $\Phi_0 = 2.0678 \times 10^9$ Oe Å$^2$ is a flux quantum.[29] This relation gives $\xi(0) = 43.6$ Å and 48.4 Å for $H_{c2}(0) = 17.3$ T and 14.04 T, respectively. Similarly, from the relation $H_{c1}(0) = (\Phi_0/4\pi\lambda^2)\ln(\lambda/\xi)$, we find the magnetic penetration depth $\lambda(0) = 3625$ Å. The Ginzburg-Landau parameter is then $\kappa = \lambda/\xi = 83.14$. Using these parameters and the relation $H_{c2}(0) H_{c1}(0) = H_c(0)^2[\ln\kappa(0) + 0.08]$ the thermodynamic critical field $H_c(0)$ was found to be 0.13 T.[34] This value of $H_c(0)$ estimates slightly smaller value of the condensation energy, $U(0)$, than that is calculated from $\gamma$.

Assuming the g-factor for the conduction electrons is close to 2, we estimate the Pauli limiting field for Nb$_{.18}$Re$_{.82}$ from $\mu_0 H^{Pauli} = \Delta(0)/\mu_B\sqrt{2}$, to be 16.8 T. This value of $\mu_0 H^{Pauli}$ is close to the estimated orbital field $\mu_0 H_{c2}(0) \approx 14.04$ T. It closely follows the relation $H_{c2}(0)/H^P = \alpha/\sqrt{2}$. Maki parameter, $\alpha$, can be derived from the expression $\alpha = 5.3 \times 10^{-5}$ $[-dH_{c2}/dT]_{T=Tc}$, revealing $\alpha = 1.2$.[35] The sizable Maki parameter obtained from this approximation is an indication that Pauli limiting is non-negligible. It is possible that due to different masses of Nb and Re the substitutional disorder has a substantial spin-orbit component. If present, this scattering suppresses the efficiency of paramagnetic limiting, and restores the upper critical field to its orbital value in a dirty conventional s-wave system. It is possible that the orbital upper critical field is enhanced due to the interplay of inversion symmetry breaking with disorder.[36] It is highly desirable to extend the investigations of the upper critical field to lower temperatures to determine its detailed behavior. If the upper critical field in the clean single crystal exceeds the Pauli limit, it would suggest a substantial contribution from the triplet component to the pairing amplitude, and other effects predominantly due to the broken inversion symmetry. Further studies of the anisotropy of the critical field through resistivity and specific heat measurements in single crystals would be even more enlightening, since they would shed light on the structure of the spin-orbit coupling in momentum space [16]. The experimentally determined values of all parameters of our sample and other reference compounds are listed in the Table 1.

## IV. CONCLUSION

We prepared polycrystalline samples of Nb$_{0.18}$Re$_{0.82}$ by arc-melting techniques. The results of X-ray diffraction measurements confirm the material's noncentrosymmetric $\alpha$-Mn structure type. From resistivity, magnetic susceptibility, and specific heat measurements Nb$_{0.18}$Re$_{0.82}$ was confirmed as a type II superconductor with a bulk transition temperature near 8.8 K. Since the upper critical field $H_{c2}(0)$ is close to the estimated Pauli limiting field, further investigations into the possible mixture of single and triplet pairing states need to be done. The high value of $\Delta C/\gamma T_c$ ~ 1.86 indicates that the compound is in the moderate-coupling regime. This is supported by the fairly large normal state electronic specific heat, $\gamma = 53.4$ mJ/mole K$^2$ and the electron phonon coupling constant, $\lambda_{ep} = 0.73$. The low temperature specific heat ($C_{es}$) behavior suggests classifying Nb$_{0.18}$Re$_{0.82}$ as a conventional BCS-type superconductor with a full gap energy of $\Delta(0) = 31.2$ K. At low temperature our best fit to the experimental data is exponential, but impurity effects or disorder in the polycrystalline sample may have masked the underlying gap anisotropy and the admixture of the triplet pairing component due to lack of inversion symmetry. The availability of single crystals is highly desirable to help answer these questions.

## V. ACKNOWLEDGMENTS

P.W.A. acknowledges the support of DOE under Grant No. DE-FG02-07ER46420, D.P.Y. acknowledges the support of the NSF under Grant No. DMR-0449022, I. V. acknowledges support from DOE Grant DE-FG02-08ER46492, and J.Y.C. acknowledges the support of the NSF under Grant No. DMR-0756281.

[1] E. Bauer, G. Hilscher, H. Michor, Ch. Paul, E.W. Scheidt, A. Gribanov, Yu. Seropegin, H. Noël, M. Sigrist, and P. Rogl, Phys. Rev. Lett. **92**, 027003 (2004).

[2] L. P. Gor'kov, and E.I Rashba, Phys. Rev. Lett. **87**, 037004 (2001).

[3] P. A. Frigeri, D.F.Agterberg, A. Koga, and M. Sigrist, Phys. Rev. Lett. **92**, 097001 (2004).

[4] M. Sigrist, D. F. Agterberg, P. A. Frigeri, N. Hayashi, R. P. Kaur, A. Koga, I. Milat, K. Wakabayashi, and Y. Yanase, J. Magn. and Magn. Mater. **310**, 536 (2007).

[5] P.A. Frigeri, D. F. Agterberg, I. Milat, and M. Sigrist, Eur. Phys. J. B **54**, 435 (2006).

[6] P. A. Frigeri, D. F. Agterberg, and M. Sigrist, New J. Phys. **6**, 115 (2004).

[7] H.Q. Yuan, D. F. Agterberg, N. Hayashi, P. Badica, D. Vandervelde, K. Togano, M. Sigrist, and M. B. Salamon, Phys. Rev. Lett. **97**, 017006 (2006).

[8] N. Kimura, K. Ito, K. Saitoh, Y. Umeda, H. Aoki, and T. Terashima, Phys. Rev. Lett. **95**, 247004 (2005).

[9] I. Sugitani, Y. Okuda, H. Shishido, T. Yamada, A. Thamizhavel, E. Yamamoto, T.D. Matsuda, Y. Haga, T. Takeuchi, R. Settai, and Y. Onuki, J. Phys. Soc. Jpn. **75**, 043703 (2006).

[10] T. Akazawa, H. Hidaka, T. Fujiwara, T. C. Kobayashi, E. Yamamoto, Y. Haga, R. Settai, and Y. Onuki, J. Phys.: Condens. Matter **16,** L29–L32 (2004).

[11] M. Yogi, Y. Kitaoka, S. Hashimoto, T. Yasuda, R. Settai, T. D. Matsuda, Y. Haga, Y. Ōnuki, P. Rogl, and E. Bauer, Phys. Rev. Lett. **93**, 027003 (2004).

[12] K. Izawa, Y. Kasahara, Y. Matsuda, K. Behnia, T. Yasuda, R. Settai, and Y. Onuki, Phys. Rev. Lett. **94**, 197002 (2005).

[13] K. Togano, P. Badica, Y. Nakamori, S. Orimo, H. Takeya, and K. Hirata, Phys. Rev. Lett. **93**, 247004 (2004).

[14] P. Badica, T. Kondo, and K. Togano, J. Phys. Soc. Jpn. **74**, 1014 (2005).

[15] T. Klimczuk, Q. Xu, E. Morosan, J. D. Thompson, H. W. Zandbergen, and R. J. Cava, Phys. Rev. B **74**, 220502 (2006).

[16] T. Shibayama, M. Nohara, H. Aruge Katori, Y. Okamoto, Z. Hiroi, and H. Takagi, J. Phys. Soc. Jpn. **76**, 073708 (2007).

[17] T. Klimczuk, F. Ronning, V. Sidorov, R. J. Cava, and J. D. Thompson, Phys. Rev. Lett. **99**, 257004 (2007).

[18] A. B. Karki, Y. M. Xiong, I. Vekhter, D. Browne, P. W. Adams, D. P. Young, K. R. Thomas, Julia Y. Chan, H. Kim, and R. Prozorov, Phys. Rev. B **82**, 064512 (2010).

[19] E. Bauer, G. Rogl, Xing-Qiu Chen, R. T. Khan, H. Michor, G. Hilscher, E. Royanian, K. Kumagai, D. Z. Li, Y. Y. Li, R. Podloucky, and P. Rogl, Phys. Rev. B **82**, 064511 (2010).

[20] M. Nishiyama, Y. Inada, and Guo-qing Zheng, Phys. Rev. Lett. **98**, 047002 (2007).

[21] K. Tahara, Z. Li, H. X. Yang, J. L. Luo, S. Kawasaki, and Guo-qing Zheng, Phys. Rev. B **80**, 060503(R) (2009).

[22] H. Takeya, M. ElMassalami, S. Kasahara, and K. Hirata, Phys. Rev. B **76**, 104506 (2007).

[23] E. M. Savitskii, V. V. Baron, Yu. V. Efimov, M. I. Bychkova, and L. F. Myzenkova, *Superconducting Materials*, Plenum Press (New York- London, 1966).

[24] R. Steadman and P.M. Nuttall, Acta Cryst. **17**, 62 (1964).

[25] N.R. Werthamer, E. Helfand, and P.C. Hohenberg, Phys. Rev. **147**, 295 (1966).

[26] H. Padamsee, J. E. Neighbor, and C. A. Shiffman, J. Low Temp. Phys. **12**, 387 (1973)

[27] H. Takeya, K. Hirata, K. Yamaura, K. Togana, M. EIMassalami, R. Rapp, F. A. Chaves, B. Ouladdiaf, Phys. Rev. B **72**, 104506 (2005).

[28] P.B. Allen and R. C. Dynes, Phys. Rev. B **12**, 905 (1975).

[29] J. P. Carbotte, Rev. Mod. Phys. **62**, 1027 (1990).

[30] C. Kittle, *Introduction to Solid State Physics*, 4$^{th}$ ed. (Wiley, New York, 1966).

[31] K. G. Wilson, Rev. Mod. Phys. **47**, 773 (1975).

[32] K. Yamada, Prog. Theor. Phys. **53**, 970 (1975).

[33] N. Tsujii, K. Yoshimura, and K. Kosuge, J. Phys.: Condens. Matter **15**, 1993 (2003).

[34] A. Junod, in *Studies of High Temperature Superconductors*, edited by A. Norliker (Nova Science, NewYork; 1996) Vol. 19.

[35] K. Maki, Phys. Rev. **148**, 362 (1966).

[36] K. V. Samokhin, arXiv:1002.4639v1 (2010)

[37] *Handbook of Superconductivity* edited by C.P. Poole, Jr. (Academic Press, New York, 1999) Ch.9, Sec. G, 478

Table I. Superconducting and other physical parameters of $Nb_{0.18}Re_{20.82}$

| Parameters | Unit | $Nb_{0.18}Re_{0.82}$ Sample | $Mo_3Al_2C$ [Ref 18] | $Mg_{10}Ir_{19}B_{16}$ [Ref 17] |
|---|---|---|---|---|
| $T_c$ | K | 8.8 | 9.2 | 4.45 |
| $\mu_0H_{c1}(0)$ | T | 0.0055 | 0.0047 | 0.0030 |
| $\rho_0$ | $\mu\Omega$ cm | 189 | 177.25 | |
| $\mu_0H_c(0)$ | T | 0.13 | 0.14 | 0.028 |
| $\mu_0H_{c2}(0)$ | T | 17.3 | 18.2 | 0.77 |
| $\xi(0)$ | Å | 43.6 | 42.3 | 206 |
| $\lambda(0)$ | Å | 3625 | 3755 | 4040 |
| $\kappa(0)$ | | 83.14 | 88.56 | 20 |
| $\gamma(0)$ | mJ/mol K$^2$ | 53.4 | 18.65 | 52.6 |
| $\Delta C/\gamma T_c$ | | 1.86 | 2.14 | 1.60 |
| $\mu_0H^{pauli}$ | T | 16.8 | 19 | 8.2 |
| $2\Delta_0$ | K | 31.9 | 36.18 | 16 |
| $\mu_0(dH_{c2}/dT)_{T=T_c}$ | T/K | -2.31 | -2.39 | -0.270 |
| $\Theta_d$ | K | 98.98 | 338.52 | 280 |
| $\lambda_{e, ph}$ | | 0.73 | | 0.66 |

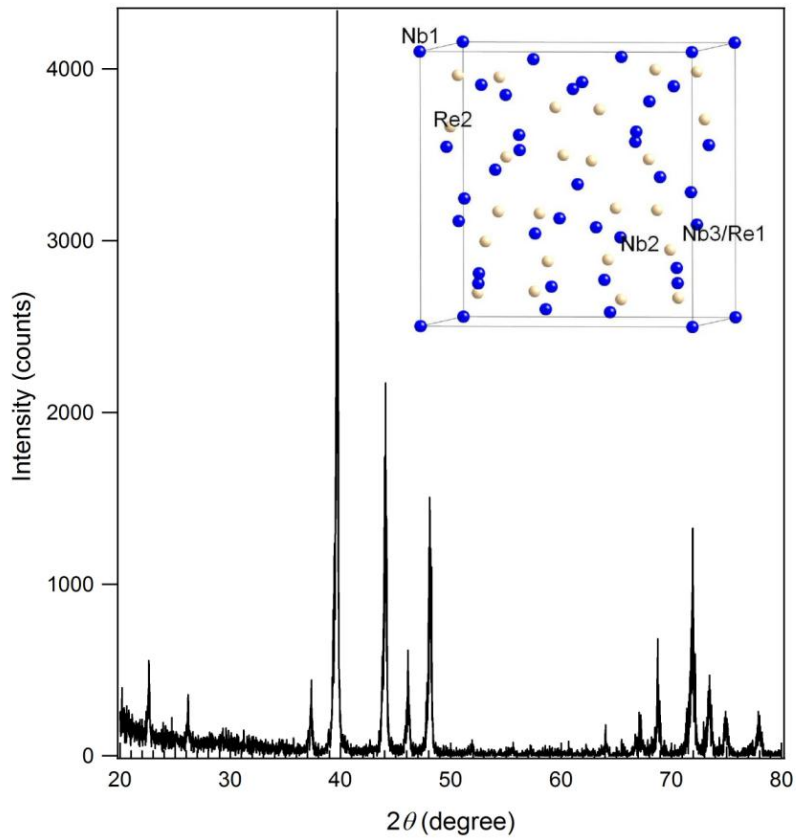

Fig. 1. Powder X-ray diffraction pattern and schematic view of the crystal structure of polycrystalline $Nb_{0.18}Re_{0.82}$.

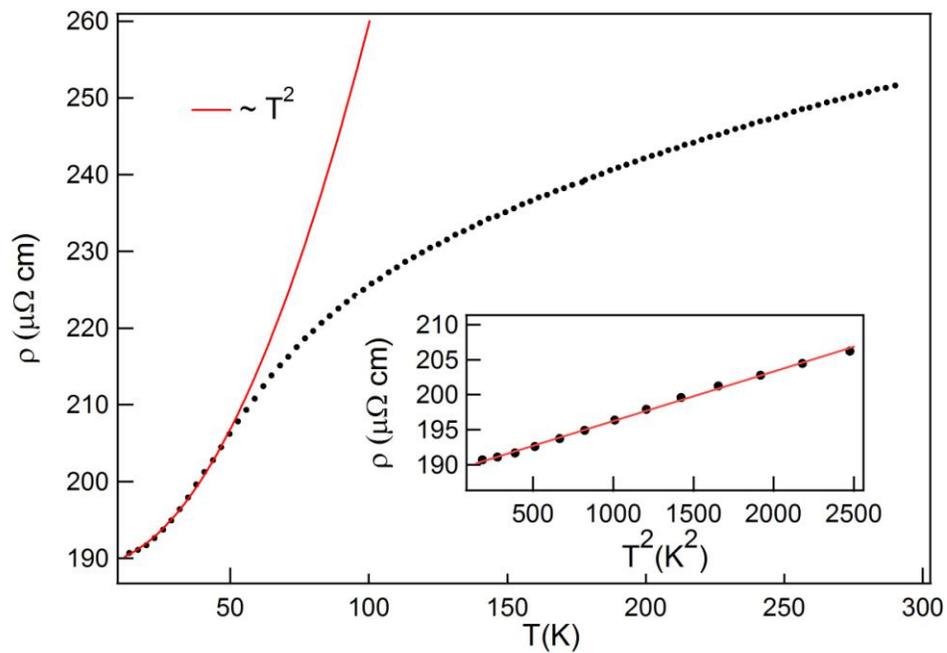

Fig. 2. The temperature dependence of the normal state resistivity for a sample of $Nb_{0.18}Re_{0.82}$. The solid lines are fits to $\rho(T)$ as indicated in the text.

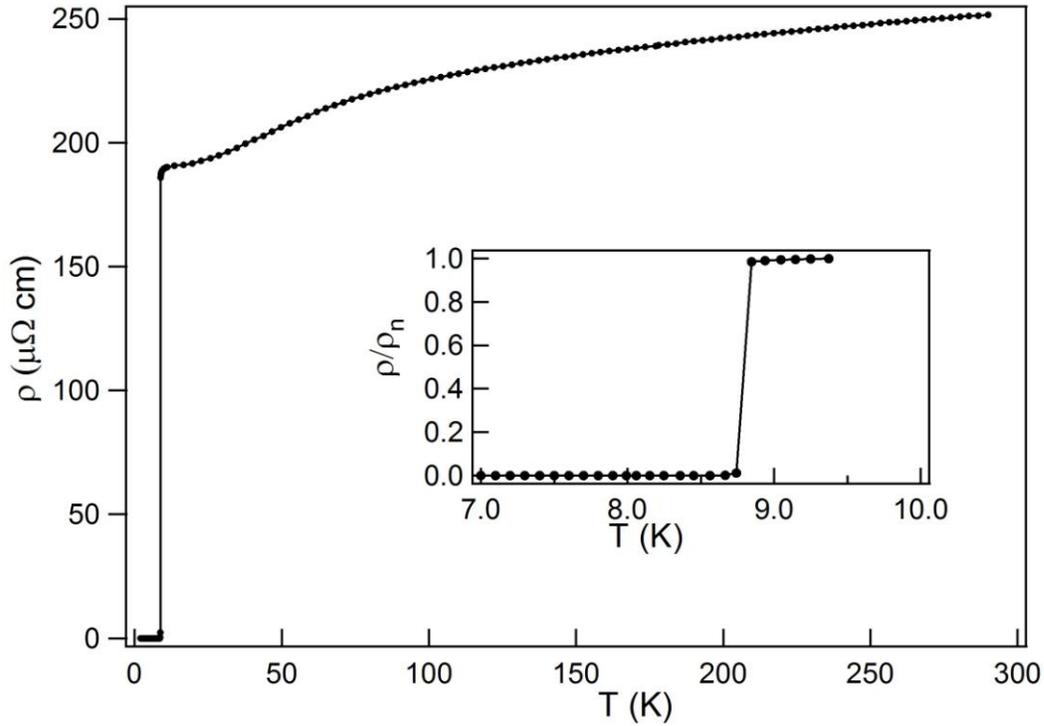

Figure 3. Temperature dependence of the resistivity of polycrystalline $Nb_{.18}Re_{.82}$ over the range of temperature from 2 to 290 K. The inset shows the superconducting transition at low temperature. The solid lines in both figures are a guide to the eye.

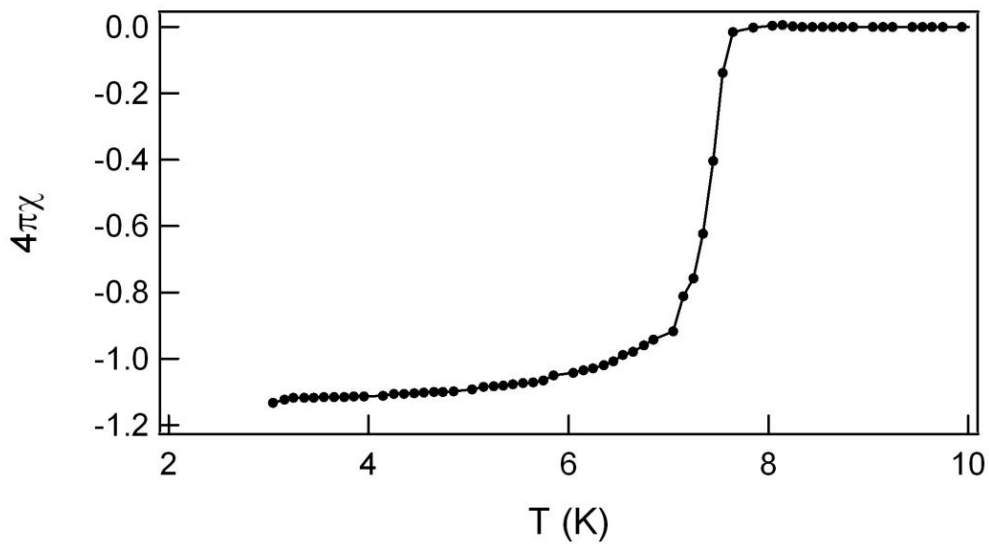

Fig. 4. ZFC magnetic susceptibility measured at constant magnetic field of 15 Oe.

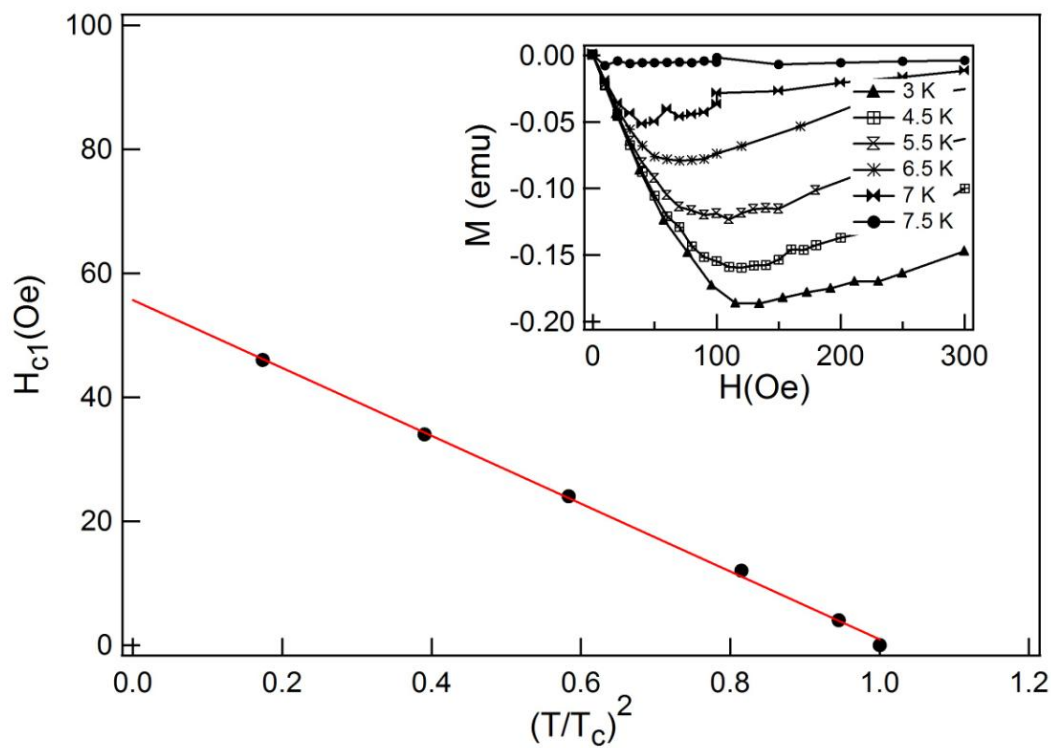

Fig. 5. Lower critical field determined by the magnetization measurements. The solid line is the best fit to the Eq. (2). Inset: Magnetization data with applied magnetic field at various temperatures.

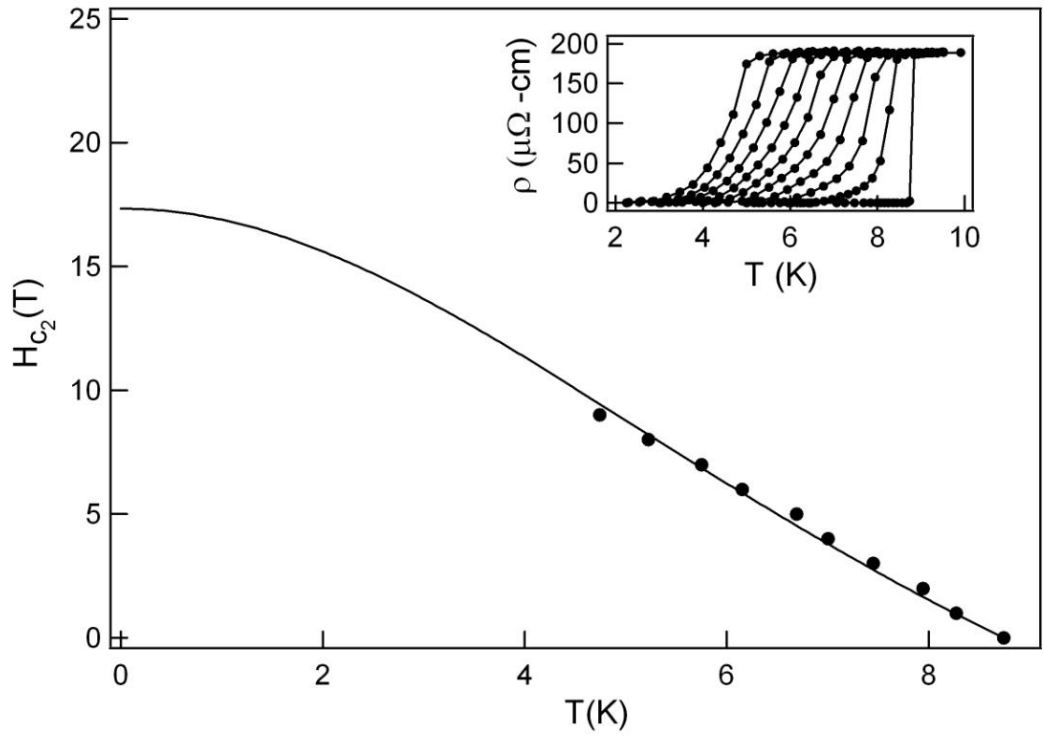

Fig. 6. Upper critical field of $Nb_{0.18}Re_{0.82}$ as a function of temperature. The solid line is fit to Ginzburg-Landau theory (see text for details). Inset: Characterization of the superconducting transition under magnetic fields 0 T to 9 T from right to left.

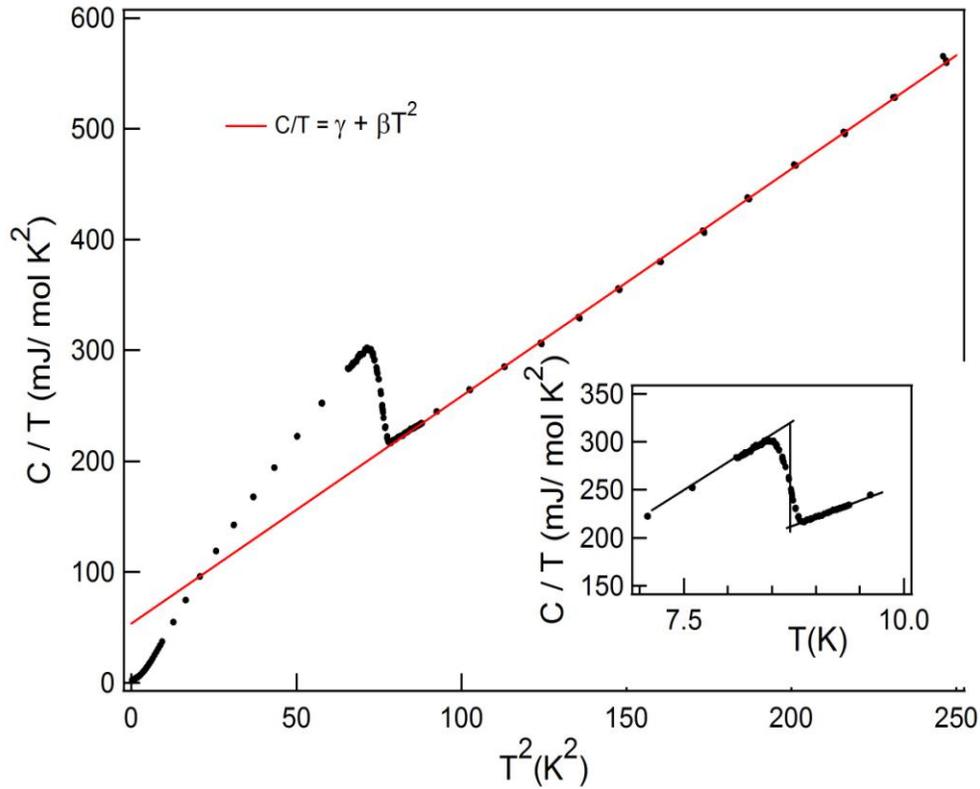

Fig. 7. Specific heat of $Nb_{0.18}Re_{0.82}$. Inset: $C/T$ vs. $T$ plot to determine the specific heat jump at the superconducting transition.

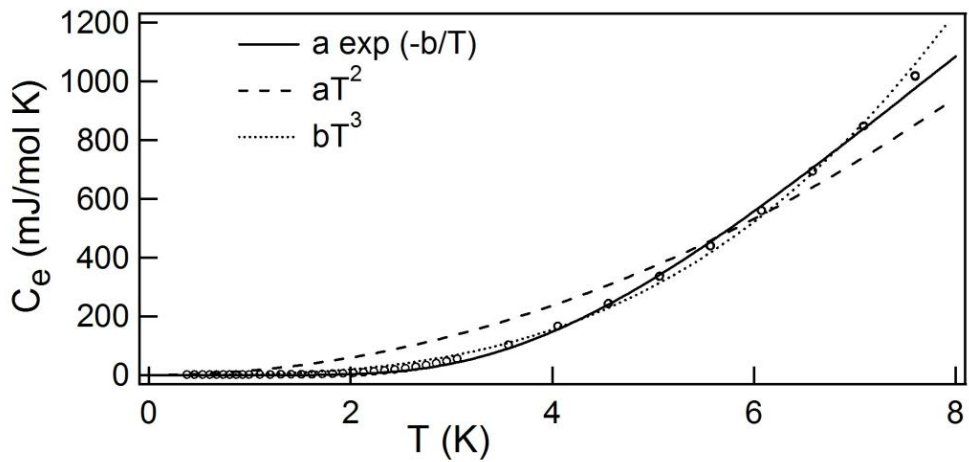

Fig. 8. Temperature dependence of electronic specific heat $C_e$ of $Nb_{0.18}Re_{0.82}$ in the superconducting phase.